\documentclass[aip,jcp,citeautoscript,citeautoscript,showkeys,floatfix]{revtex4-1}
  \usepackage{natbib}
\usepackage{graphicx}
\usepackage{float}
\usepackage{amsmath}
\usepackage{wasysym}
\usepackage{textcomp}
\usepackage[version=3]{mhchem}
\usepackage[linktocpage,pdfstartview=FitH,colorlinks,linkcolor=blue,anchorcolor=blue,citecolor=blue,filecolor=blue,menucolor=blue,urlcolor=blue]{hyperref}
\date{\today}
\title{}
\begin{document}

\raggedbottom

\title{Machine-learning accelerated geometry optimization in molecular simulation}

\author{Yilin Yang}
\affiliation{Department of Chemical Engineering, Carnegie Mellon University, 5000 Forbes Ave, Pittsburgh, PA 15213}

\author{Omar A. Jiménez-Negrón}
\affiliation{Department of Chemical Engineering, University of Puerto Rico-Mayagüez, Mayagüez, PR 00681, Puerto Rico, USA}

\author{John R. Kitchin}
\email{jkitchin@andrew.cmu.edu}
\affiliation{Department of Chemical Engineering, Carnegie Mellon University, 5000 Forbes Ave, Pittsburgh, PA 15213}

\date{\today}

\begin{abstract}
Geometry optimization is an important part of both computational materials and surface science because it is the path to finding ground state atomic structures and reaction pathways. These properties are used in the estimation of thermodynamic and kinetic properties of molecular and crystal structures. This process is slow at the quantum level of theory because it involves an iterative calculation of forces using quantum chemical codes such as density functional theory (DFT), which are computationally expensive, and which limit the speed of the optimization algorithms. It would be highly advantageous to accelerate this process because then one could either do the same amount of work in less time, or more work in the same time. In this work, we provide a neural network (NN) ensemble based active learning method to accelerate the local geometry optimization for multiple configurations simultaneously. We illustrate the acceleration on several case studies including bare metal surfaces, surfaces with adsorbates, and nudged elastic band (NEB) for two reactions. In all cases the accelerated method requires fewer DFT calculations than the standard method. In addition, we provide an ASE-optimizer Python package to make the usage of the NN ensemble active learning for geometry optimization easier.
\end{abstract}

\maketitle

\section{Introduction}
\label{sec:org666236c}

Machine learning has been reshaping the research methods of many scientific and engineering fields. In the area of surface catalysis, various applications of machine learning techniques are emerging that enable larger simulations of nanoparticles  \cite{jinnouchi-2017-predic-catal}, structure optimization \cite{jacobsen-2018-fly-machin,hansen-2019-atomis-machin}, studies of segregation \cite{boes-2017-model-segreg}, high throughput screening \cite{lamoureux-2019-machin-learn,back-2019-convol-neural} and on the fly learning of force fields  \cite{vandermause-2020-fly-activ}. One of the crucial requirements for a machine learning model to work is a broad training dataset which ensures the generalization ability of complex machine learning model on the test dataset. For example, accurate adsorption energies of certain adsorbates on various kinds of catalytic surfaces is one of the basic prerequisites to conduct high-throughput screening for novel catalyst candidates \cite{li-2017-high-throug,ling-2018-gener-two,zhang-2020-amorp-catal}. Thus, many studies aim to build up a reliable machine learning model to predict the adsorption energies on different adsorption sites \cite{gu-2012-shape-contr,ulissi-2017-machin-learn,hoyt-2019-machin-learn}. In this case, a training set covering most of the possible configurations is necessary to obtain a reasonable model which affects the reliability of the screening process.

The rate-limiting step to obtain the adsorption energies is often the geometry optimization process. This process consists of a sequence of iterative single point calculations with density functional theory (DFT), with the structure update is completed by various optimizers like conjugate gradient descent or the Broyden–Fletcher–Goldfarb–Shanno (BFGS) algorithms. These algorithms start with an initial guess, and then iteratively move the atoms to reduce the forces to a specified tolerance. The forces are typically computed at each step by the DFT code. One path to speeding up these calculations is to use a better initial guess. An alternative approach is to use a surrogate model that is computationally cheap, but sufficiently accurate that many steps can be taken with the cheap model before a DFT calculation is required. Recently, many machine learning methods have been developed to accelerate the local geometry optimization process with this idea. For example, Peterson \cite{peterson-2016-accel-saddl} used a neural network as the surrogate model to find the transition state, but the uncertainty is not included. Torres et al \cite{torres-2019-low-scalin} and Koistinen et al \cite{koistinen-2017-nudged-elast} used Gaussian Process Regression (GPR) to estimate the uncertainty during the local geometry optimization. Those implementations of GPR are solely based on the Cartesian coordinates of the atoms, which limits the training set to the past geometries of the same configuration size and composition during the optimization and the information of other configurations can not be utilized. There are other applications of active learning in geometry optimization \cite{artrith-2012-high-dimen,rio-2019-local-bayes,rio-2020-machin-learn,vandermause-2020-fly-activ,shuaibi-2021-enabl-robus} or in molecular dynamics \cite{tong-2018-accel-calyp,jinnouchi-2020-fly-activ}.  Most of these methods are also based on active learning with uncertainty measured by Gaussian process regression or a neural network (NN) ensemble. In the active learning relaxation process, a surrogate model is trained to replace the expensive DFT calculation to conduct the energy minimization steps. At each step the uncertainty of the model prediction is monitored. If the uncertainty exceeds a specified threshold, DFT calls will be requested to get accurate energy and force information for the uncertain configuration. Then this new data point is used to update the surrogate model to improve it.

The work to date has mostly focused on the relaxation of a single configuration, which might have limited acceleration when applied to relax many configurations. In each case, the surrogate model essentially starts from scratch, and has no ability to share information between similar configurations. In this work, we illustrate and evaluate an online learning method to accelerate the local geometry optimization for multiple configurations simultaneously. More specifically, we focus on two aspects to accelerate the online learning process. The first point is the training of the surrogate model used to relax the target configurations. When the training set grows up, the training of the machine learning model also takes more time, which might resulted in longer relaxation time than using DFT solely, although with less number of DFT calls. This issue is shared among various machine learning models including GPR and deep learning model. We note that using local training dataset is sufficient to conduct the local geometry optimization at each step. Thus, the size of the training set used to update the surrogate model at each step could be limited, which could significantly reduce the time used to train the surrogate models. The second point of this work is to discuss the potential methods that could be adapted to accelerate the active learning relaxation process for large number of configurations. We illustrate three adaptations to three different scenarios: relaxation from scratch, relaxation from a small dataset and relaxation from a large existing dataset. The main point under these methods is that the information of different relaxation trajectories could be shared among each other to accelerate the overall relaxation process. Another objective of this work is to provide an overview about the performance of NN-based online learning on various local geometry optimization tasks.

\section{Methodology}
\label{sec:orgc208727}

\subsection{Machine Learning Model for the surrogate models}
\label{sec:orgfa5dc64}

Many machine learning models have been established to model the potential energy surface (PES) such as the Gaussian Approximation Potentials (GAP) \cite{bartok-2010-gauss-approx-poten} and Behler Parrinello Neural Networks (BPNN) \cite{behler-2007-gener-neural}. In this work, we choose the single neural network (SingleNN) as our basic model to approximate the energies and forces information of the atomic configurations \cite{liu-2020-singl}. SingleNN is a modified BPNN which represents different elements by multiple output nodes in a single NN instead of separate NNs. SingleNN uses the same symmetry functions as the conventional BPNN, but uses a single neural network with an output for each element, rather than a separate neural network for each element. Under the same NN structure (same number of hidden layers and same nodes in each layer), it contains fewer parameters than BPNN. Thus, the training and inference time is lower. The feature representation of the atomic environment used in this work is the atom-centered symmetry function (ACSF) \cite{behler-2011-atom-center}. The NN structure contains two hidden layers with 50 neurons at each layer. The activation function used is tanh. These hyperparameters were chosen by cross validation among different NN architectures on the dataset of previous work and they are typical for machine learned potentials. The structure of this NN looks relatively over-parameterized considering the small size of the dataset in this work (typically the dataset contains 50 configurations). This is because we want to utilize the benefits of an over-parameterized deep learning model: 1) With high probability, convergence of the training process is easy from a random initialization, and 2) an over-parameterized NN could lead to unsimilar models with high probability only using different random initializations \cite{NEURIPS2019_62dad6e2,lakshminarayanan-2016-simpl-scalab}. There is also some applications of large NN models on small datasest with reasonable generalization ability \cite{NEURIPS2018_fface838}. We used early stopping to prevent the overfitting of the NN model. For our specific application, we need to note that overfitting is not expected to be a severe problem because 1) the surrogate model is only used when uncertainty is low, 2) if the uncertainty exceeds a threshold the data is augmented by new DFT data, and 3) the final minimum is always validated by DFT.

The atomic energy, total energy and forces predicted by our SingleNN could be formulated by Equation \ref{snn-predict1} - \ref{snn-predict4}.

\begin{equation}
\label{snn-predict1}
\textbf{o}_{i} = \textbf{W}^{(2)} f_{a}^{(2)} \left( \textbf{W}^{(1)} f_{a}^{(1)} \left( \textbf{W}^{(0)} \textbf{g}_i + \textbf{b}^{(0)} \right) + \textbf{b}^{(1)} \right) + \textbf{b}^{(2)}
\end{equation}

\begin{equation}
\label{snn-predict2}
E_{i} = mask_{i} \cdot \textbf{o}_{i}
\end{equation}

\begin{equation}
\label{snn-predict3}
E_{tot} = \sum_{i}^{N} E_{i}
\end{equation}

\begin{equation}
\label{snn-predict4}
\textbf{f}_{i} = - \frac{\partial E_{tot}}{\partial \textbf{r}_{i}}
\end{equation}

In these equations \(\textbf{o}_{i}\) is the output layer of the SingleNN for atom \(i\), \(\textbf{g}_i\) is the fingerprint vector for atom \(i\), \(f_{a}^{(l)}\), \(\textbf{W}^{l}\), \(\textbf{b}^{(l)}\) are the activation function, weight and bias at layer \(l\). \(mask_{i}\) is a one-hot vector indicating the element of the atom \(i\).  \(E_{i}\), \(\textbf{f}_{i}\) are the energy and forces of atom \(i\). \(N\) is the number of atoms in a configuration. \(E_{tot}\) is the total energy of the configuration.

To measure the uncertainties of the model predictions, we adopt the NN ensemble method as an approximate estimation \cite{lakshminarayanan-2016-simpl-scalab}. We use 10 NNs in the NN ensemble and each NN has the same structure. As mentioned in the original ensemble method paper, each NN is trained on the same training set without bootstrapping but with different random initialization. This is because different initializations are already able to generate different NN models using the same training set because of the over-parameterization of the NN model. \cite{NEURIPS2019_62dad6e2}

The prediction uncertainty is estimated by the variance of the model predictions in the ensemble. We used a relative ratio to the maximum variance of the NN ensemble in the training set as a criterion to check if a configuration is uncertain or not. More specifically, Equation \ref{threshold} quantifies this uncertainty threshold.

\begin{equation}
\label{threshold}
thld = \alpha  \max_{i} {\mathrm{Var} \left[ E_{tot}^{i} \right]}
\end{equation}

\begin{figure}[htbp]
\centering
\includegraphics[width=.9\linewidth]{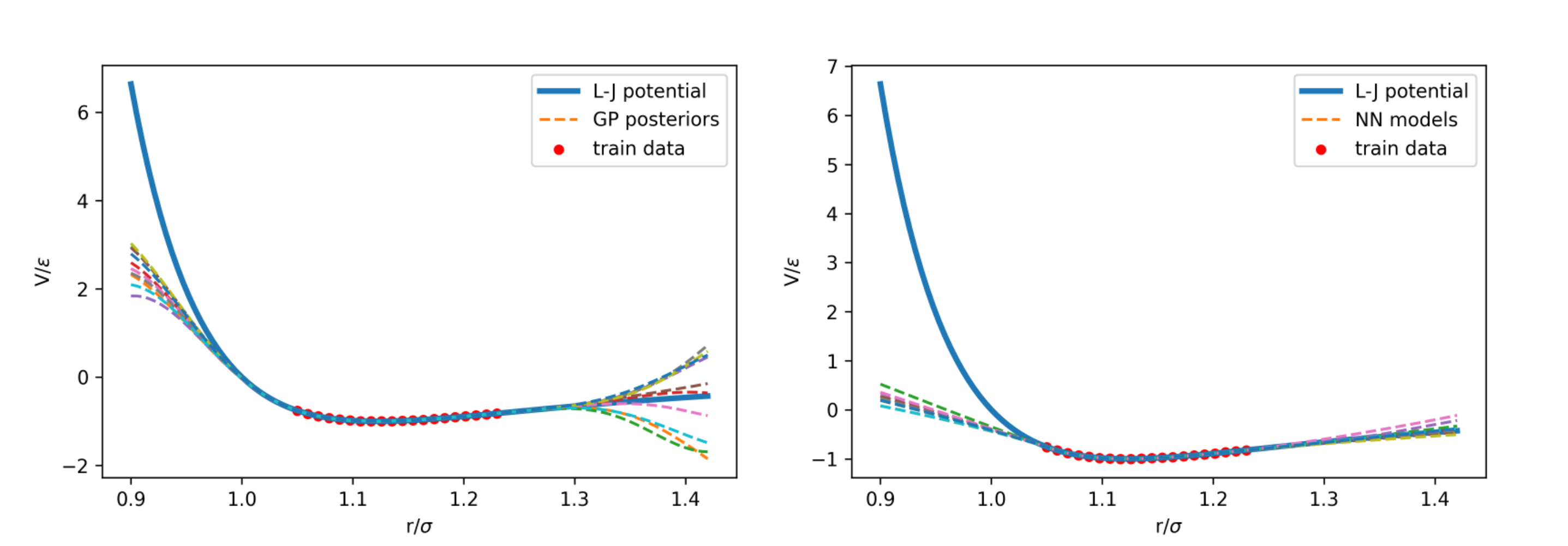}
\caption{\label{lj}Surrogate machine learning models for the Lennard Jones potential. Left plot shows the GPR while the right plot shows the NN ensemble. Both models have low prediction variance in the region of training set and high variance for the data that is far from the training set.}
\end{figure}

Where \(\alpha\) is the coefficient to control the extent to believe the prediction of the NN ensemble. \(\mathrm{Var} \left[ E_{tot}^{i} \right]\) is the prediction variance of the NN ensemble on the total energy of a configuration \(i\) in the training set. \(thld\) is the threshold above which a prediction is considered as uncertain. We chose the \(\alpha\) by comparing the performance of different values on a small dataset. For the various applications below, setting alpha between 2 to 3 works for all examples and we use 2 as the default value in the GitHub package. The intuition is that if the NN ensemble has a similar variance on a test configuration as the variance in the training set, then we could expect the test configuration is close to the region of the training dataset, thereby, we could expect similar error with the training error. If it is far away from the maximum variance in the training set, it is probable that extrapolation is occurring, and we should be careful about the prediction. This intuition is shared by different machine learning models like the GPR and the NN ensemble. For example, Figure \ref{lj} shows the GPR and NN models for the Lennard Jones potential \cite{1924-deter-molec}. Both models have small prediction variance in the region of the training data. As the test data goes far away from the training set, the prediction error and variance also increase.

We also compare this NN model with the GPR model in one of our datasets. The details of the GPR formula are attached in the supporting information. Optimization of the hyperparameters like the bandwidth and the data noise term was done according to the previous literature reports \cite{koistinen-2017-nudged-elast,torres-2019-low-scalin}. The data noise in this application could be the DFT convergence error related to the factors like k points and cutoff energy.

\subsection{Relaxation with Active Learning}
\label{sec:org3eb3651}

\begin{figure}[htbp]
\centering
\includegraphics[width=.9\linewidth]{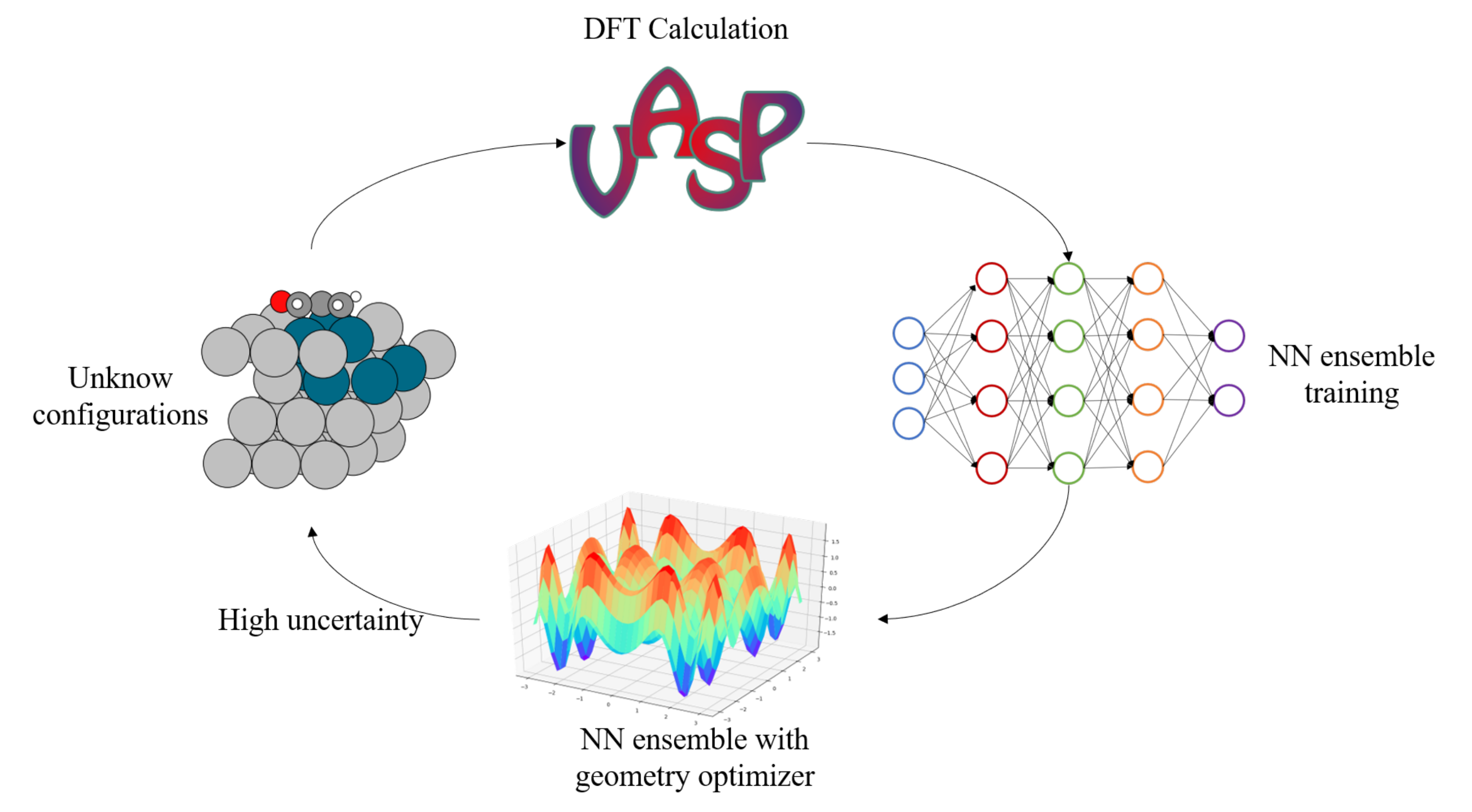}
\caption{\label{al-fmk}Framework for relaxation with online active learning. The overall workflow starts with the initial configurations that needs to be relaxed. At first, the DFT energies and forces are calculated and the NN ensemble is trained with these initial information. Then the model is utilized with optimizers to reduce the energy of the configurations. The relaxation with model stops when encounters with uncertain configurations or reaches the relaxation criterion. The uncertain configurations are submitted for further DFT calculations.}
\end{figure}

The framework of the active learning for relaxation is shown in Figure \ref{al-fmk} which is similar to most active learning frameworks, \cite{jacobsen-2018-fly-machin,vandermause-2020-fly-activ}  but we process multiple configurations simultaneously to obtain extra acceleration. The rationality of pooling different trajectories together is that the information of similar atomic environment across trajectories could be shared by a common atomic NN surrogate model, which was also observed in the water NN potential. \cite{schran-2021-trans-machin} Another benefit of the pooling is that it could be applied in a scalable way. Different configurations could share a common surrogate model and there is no need to assign separate computing resources for the training of each trajectory. For the specific procedure, we start from \(N\) configurations to be relaxed, build a common NN ensemble for these \(N\) configurations. At each step, we conduct relaxation until the model becomes uncertain for each configuration. Then we query DFT for the true energies and forces for these uncertain configurations, which are used to update the surrogate model. During the relaxation process, we limit the size of the training set and keep the configurations of the most recent steps; all previous configurations are discarded in the iterative training of the NN ensemble. This setting is used to reduce the time to train a NN when the available data points grows as the relaxation steps. Intuitively, this modification is similar to the L-BFGS compared to the BFGS, which estimates the inverse of the Hessian matrix at a point using the recent gradients instead of the full history. \cite{liu-1989-limit-memor} However, L-BFGS aims to alleviate the memory problem while we try to reduce the training time for the surrogate model.

Before running the online learning to relax the target configurations, several cases should be considered. If no prior data related to the target configurations is available, then the initial model is built on the DFT information of the initial configurations. If there are some existing relaxation trajectories that are related to the target configurations (e.g. alloys with the same elements but different configurations), then this data is incorporated with the DFT data of the initial configurations to set up the initial NN model. This part of reused data also accelerates the overall process of relaxation. Finally, if training data is available from previous relaxations that are similar to the initial configurations, then it is possible to conduct the relaxation in a offline way through the NN model trained on the prior training set without initially accessing the DFT calculation.

\subsection{Application Dataset}
\label{sec:orge84418a}

In this work, we test the proposed online learning methods on a variety of structures including bare pure metal slabs, bare metal alloy slabs, slabs with an adsorbate, and a nanoparticle with an adsorbate. These structures increase in complexity, and are expected to be increasingly expensive to do geometry optimization with. More specifically, we take Au FCC(100), Au FCC(111), Au FCC(211), Au FCC(643), Au FCC(111) with propylene on the surface, AuPd FCC(111), AgPd FCC(111) with acrolein on the surface, AuPd icosahedron  with CO on edge as the examples for these structures. For the slab, the bottom two layers are fixed and the remaining atoms are free to be relaxed. For nanoparticles, all atoms are free to move during the relaxation. In addition to the geometry relaxation of these structures, we also evaluate this method on two climbing-image nudged elastic band (CINEB) cases \cite{henkelman-2000-climb-image}: Pt heptamer rearrangement over Pt FCC(111) surface and acetylene hydrogenation over Pd FCC(111) surface. The CINEB algorithm is like a constrained geometry optimization where forces in the direction tangent to the bands are projected out. The basic framework to perform CINEB using NN ensemble is similar to the CINEB based on Gaussian Process Regression (GPR) \cite{torres-2019-low-scalin}. In our work, the surrogate model is the NN ensemble instead of the GPR. During the relaxation, when one of the configurations in the CINEB is identified in the uncertain region of the NN ensemble, we query for a DFT calculation for this configuration. This process continues until all configurations are relaxed with certainty, then we query the DFT information for the configuration with highest energy until the energy and force prediction for the highest-energy configuration is certain and the true force is lower than a specified threshold.

The DFT used in this work is performed by the Vienna Ab initio Simulation Package (VASP) \cite{kresse-1993-ab-initiom,kresse-1996-effic-ab} with Perdew-Burke-Ernzerhof generalized gradient approximation (GGA-PBE) as the exchange-correlation functional \cite{perdew-1996-gener-gradien,perdew-1997-gener-gradien}. For the Pt heptamer rearrangement case, we used EMT as the calculator for energy and forces because of the size of this system (unit cell with 343 Pt atoms) as implemented in ASE \cite{larsen-2017-atomic-simul}. The related dataset, relaxation trajectory, configurations in the NEB as well as the code used to conduct the active learning geometry optimization are available in on GitHub \cite{yang-nn}, in which the code to calculate the fingerprints is modified based on the functions of SimpleNN \cite{lee-2019-simpl-nn}.

\section{Results and Discussion}
\label{sec:org11ec788}

\subsection{Active learning for geometry optimization of single configuration}
\label{sec:orgcdb8474}

\begin{figure}[htbp]
\centering
\includegraphics[width=.9\linewidth]{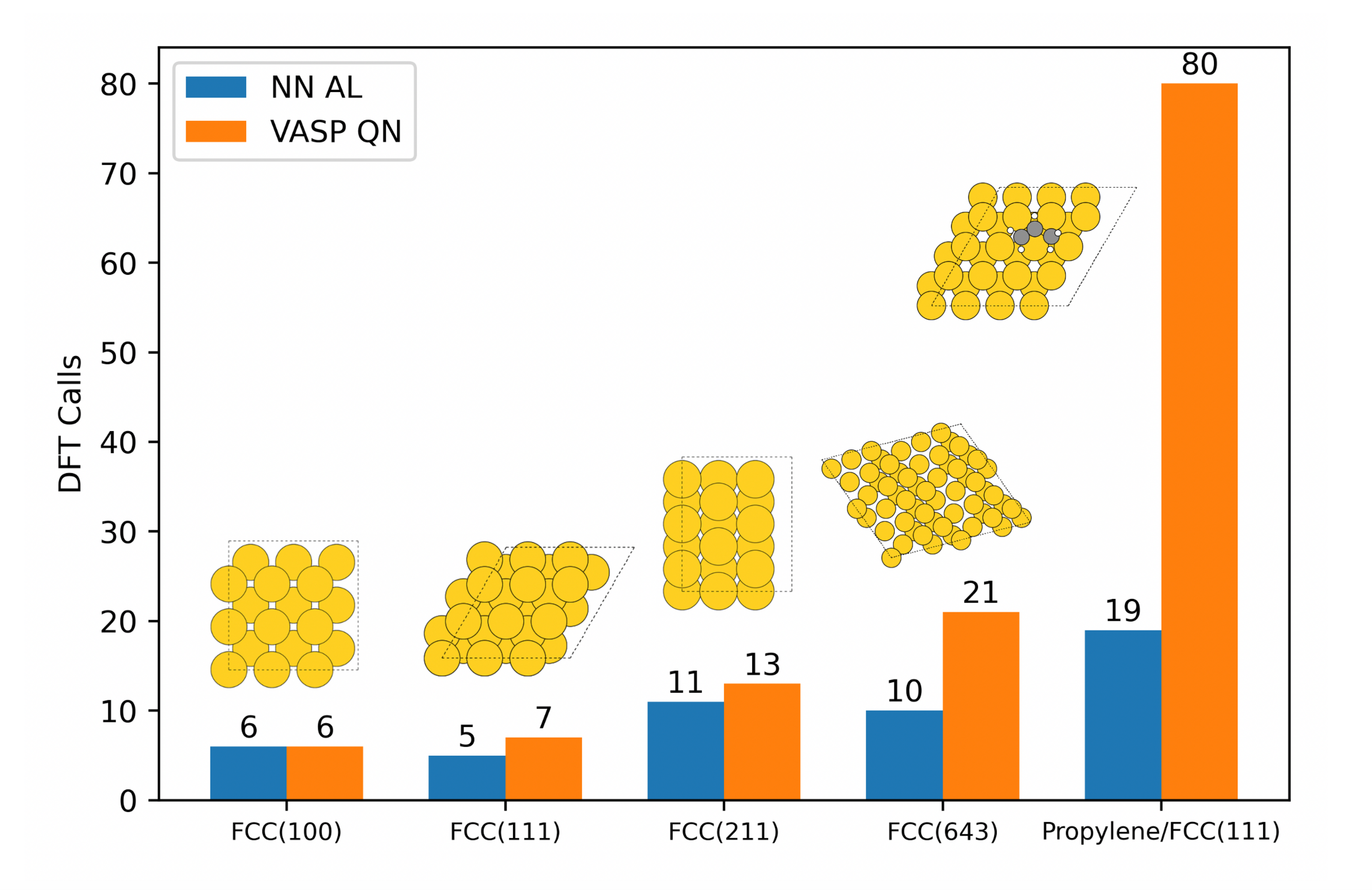}
\caption{\label{single-config}Comparison of the number of DFT calls between active learning with NN ensemble and quasi-Newton built in VASP when each configuration is relaxed independently.}
\end{figure}

Usually, geometry optimization is done for each configuration separately. For example, one may be interested in the relaxed geometry of an occupied adsorption site, then the geometry optimization would be performed on an initial guess of the configuration. Active learning could be integrated into the optimization trajectory to accelerate the process by using a surrogate model with uncertainty. With the example of Au slabs with or without an adsorbate, we evaluated the performance of active learning on single configuration relaxation and compare it with the quasi-Newton optimizer built in VASP (RMM-DIIS) \cite{pulay-1980-conver-accel}. As shown in Figure \ref{single-config}, the acceleration for the bare slabs is not as significant as it is for the slab with propylene on the top. The more complex surface FCC(643) gains more acceleration than simpler surface FCC(100), FCC(111) and FCC(211). The results suggest that the surrogate model requires a minimum number of steps or configurations to build up a sufficient approximated potential energy surface, and then to show acceleration. These results how that with active learning the number of DFT calls may be reduced by a factor of two to four for geometry optimizations that require 20 or more relaxation steps.

\subsection{Further acceleration by information sharing among configurations and utilizing prior data}
\label{sec:org726cfa9}

There are multiple ways to use machine learning to accelerate geometry optimization. First one may build the surrogate machine learned model from the relaxation trajectory of a single configuration as it develops, using the surrogate model when it is sufficiently accurate. Alternatively, one can relax many (related) configurations in parallel and train a single surrogate machine learning model on the collection of developing trajectories (the multiple method). Finally, if one has access to the relaxation trajectories from previously relaxed configurations one can pretrain the surrogate machine learning model and then use it (the warm up method).

We compare the performance of active learning with these different strategies: single configuration, multiple configurations and multiple configurations with warm up (pre-training) on the example of an adsorbed acrolein molecule on an FCC(111) alloy AgPd system. This system is more complex than the examples in the previous section with less symmetry and it is expected to take more relaxation steps to find a minimum energy geometry.  Here we use the same query strategy for new DFT single point calculations, but with different settings for the initialization. For the single configuration active learning, the method only focuses on relaxing one configuration at each time. The surrogate model starts with the DFT information of the target configuration. At each relaxation step, it relaxes this configuration and queries the DFT label for one uncertain configuration. For the multiple configurations setting, the DFT energies and forces of all target initial configurations are used to initialize the NN model. Then all configurations are optimized until each configuration is fully relaxed or goes into uncertain region of the surrogate model. In terms of the warm up setting, it requires some prior DFT data related to the target configurations that need to be relaxed, such that the surrogate model could be pre-trained with this prior DFT information which serves as the prior beliefs for the potential energy surface.

\begin{figure}[!htpb]
\centering
\includegraphics[width=.9\linewidth]{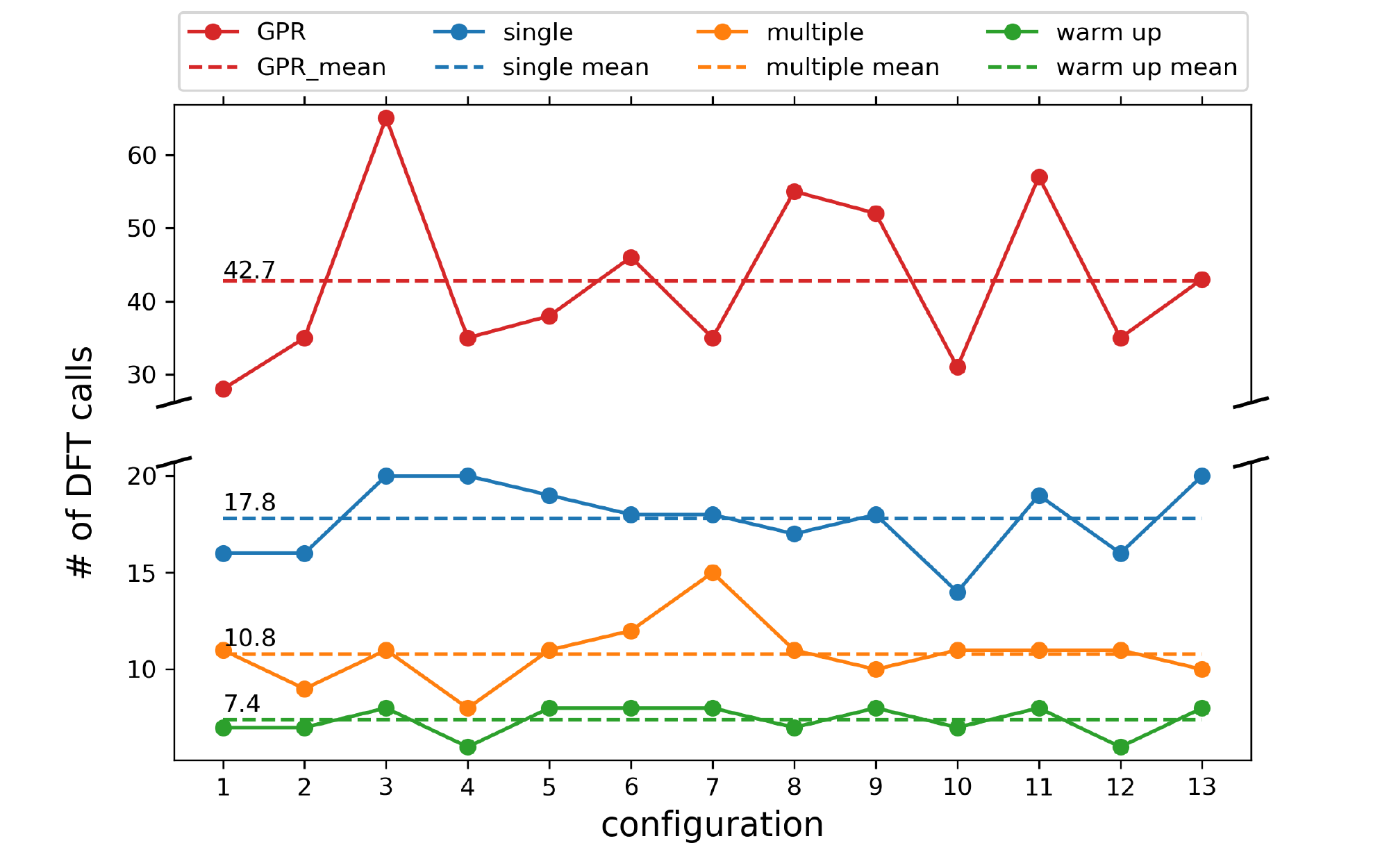}
\caption{\label{al-comp}Number of DFT calls for three different active learning setting for the relaxation of acrolein/AgPd(111). The blue line represents the single configuration mode, the orange line is for the multiple configurations mode and the green line shows the multiple configurations with warm up. The red line serves as a baseline which is the performance of GPR model implemented according to previous literatures \cite{koistinen-2017-nudged-elast,torres-2019-low-scalin}. For comparison, with no ML it takes about 193 DFT calls to converge.}
\end{figure}

The performance of above three methods on 13 different acrolein/AgPd configurations are shown in Figure \ref{al-comp}. With standard DFT/QN geometry optimization it take about 193 DFT steps on average to relax the geometries. All three methods in our work and the GPR model show acceleration, while the NN methods present better performance over the GPR model. The hyperparameters of the GPR model are referenced from the previous literature reports \cite{koistinen-2017-nudged-elast,torres-2019-low-scalin}. We note that the hyperparameters from the reported literatures might not be the optimal for our system, but even still we observe acceleration of about four times fewer steps with the GP, 11 times fewer steps for the single configuration, and thirteen times fewer steps for the multiple configurations. The pretrained warm-up shows the largest acceleration indicating that the surrogate model is more accurate and has performed better. Clearly, the information sharing through the surrogate model accelerates the active learning relaxation process. The large reduction in the number of DFT calls required directly translates to saved time and computing resources. In the limit of a fully trained machine learned potential, one can expect no additional DFT calculations are required for a new relaxation, but in our experience and in the literature it takes thousands of DFT calculations to obtain that.

\begin{figure}[!htpb]
\centering
\includegraphics[width=.9\linewidth]{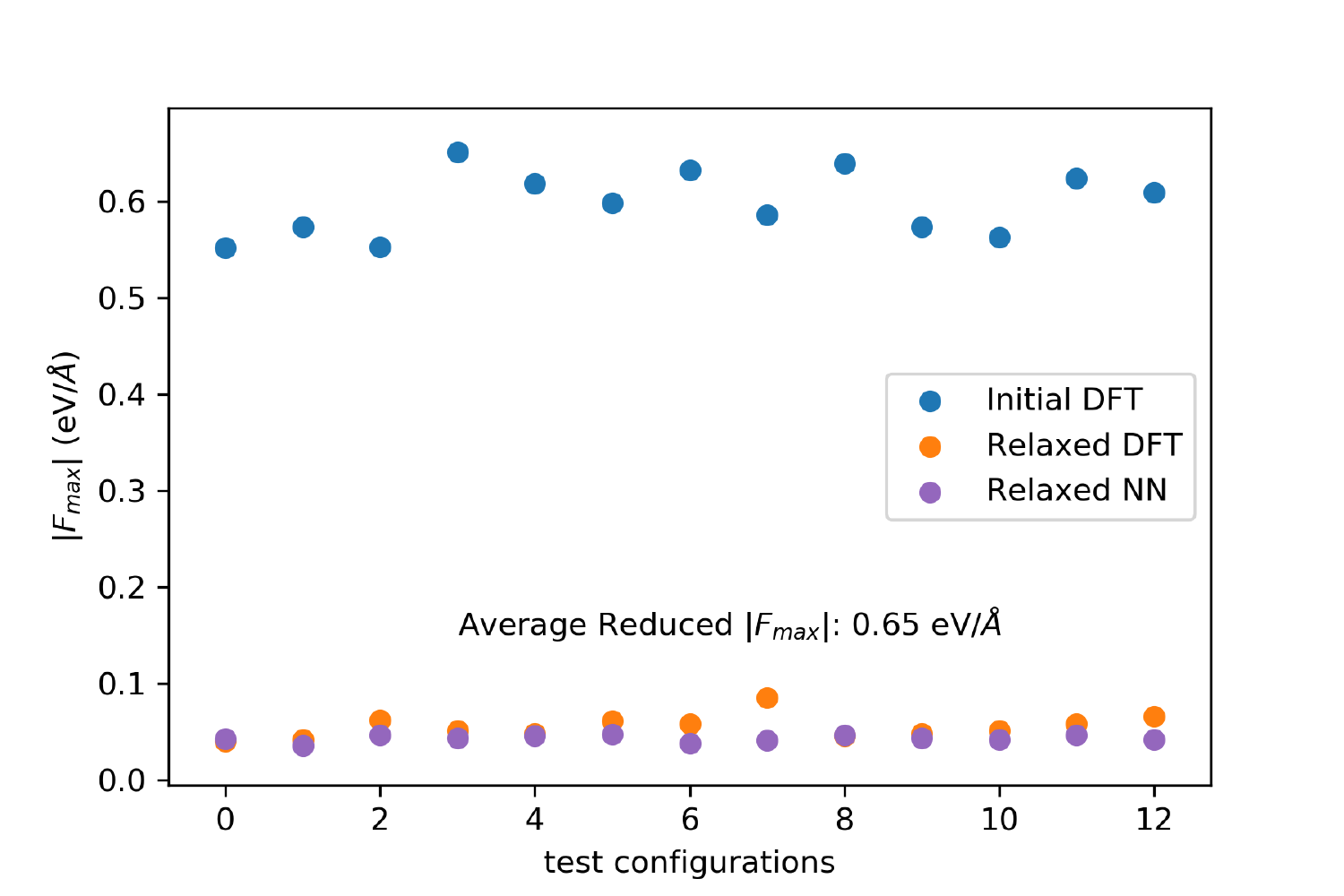}
\caption{\label{offline}Offline relaxation on 13 acrolein/AgPd configurations using NN trained on 243 existing relaxation trajectories. Blue points show the maximum DFT forces for the initial configurations. Orange scatters are the maximum DFT forces for the NN relaxed configurations while purple dots are the NN predicted maximum forces.}
\end{figure}

A related  scenario is when we have some data about the target configurations that we want to relax. For example, if we have the active learning relaxation trajectories for many configurations of acrolein/AgPd and we want to relax the remaining configurations. In this case we can utilize the existing data to build up a model to approximate the PES of the acrolein and AgPd, and then conduct the relaxation process offline since it is possible that the information required to relax the remaining configurations has been included in the existing trajectories. We show the offline relaxation performance in Figure \ref{offline}, in which 243 acrolein/AgPd relaxation trajectories are used to train a NN model. Then, another 13 configurations are relaxed using this model. Without accessing any DFT calls, the NN could reduce the maximum force of the configurations from 0.7 eV/\(\AA\) to below  0.1 eV/\(\AA\), which could serve as a preprocessing step if lower forces are required, in other words to provide better initial guesses. The NN ensembles provide uncertainty estimates, which would be useful for determining if the pretrained models are sufficiently accurate for new configurations that are not similar to the training set.

In summary, this section shows that machine learning surrogate models can be trained on the fly or in advance in a variety of ways to accelerate geometry optimization. The biggest on the fly acceleration occurs when multiple similar configurations are relaxed in parallel with shared training data in a single surrogate model. Further acceleration can be obtained if training data already exists to retrain the surrogate model on. In the next section we show the acceleration is observed for many different atomistic systems, and the degree of acceleration is system dependent.

\subsection{Performance of the active learning on more complex systems and nudged elastic band calculations}
\label{sec:org28fdb33}

\begin{figure}[htbp]
\centering
\includegraphics[width=.9\linewidth]{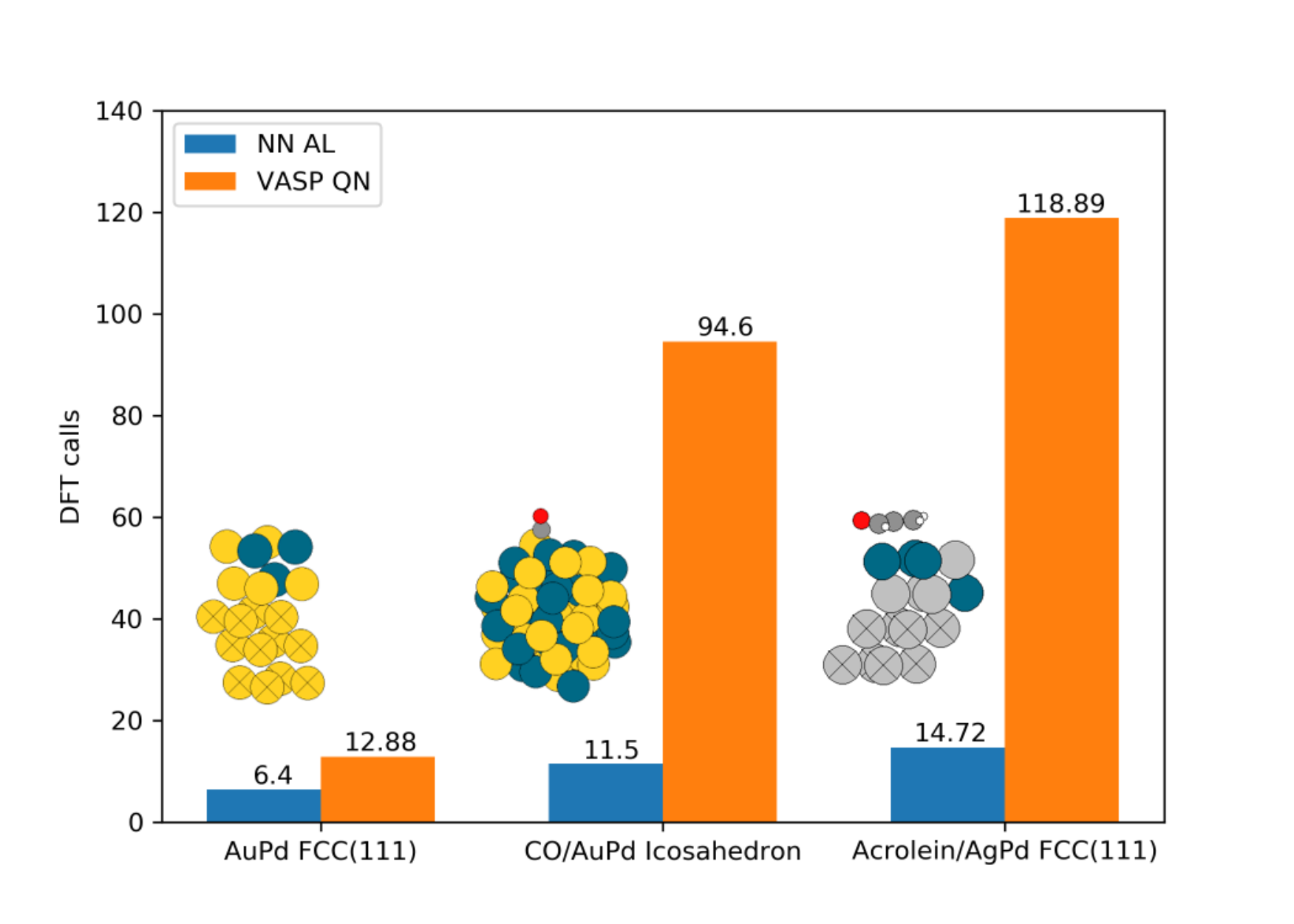}
\caption{\label{slab-nano-ads}Comparison of active learning (AL) and VASP quasi-Newton (QN) method on relaxing three different structures: bare AuPd slab, CO on AuPd icosahedron and acrolein on AgPd slab.}
\end{figure}

To explore the ability of the active learning with multiple configurations to accelerate geometry optimization, we evaluate this method on three different chemical structures: bare AuPd FCC(111) slab, CO on an AuPd icosahedron nanoparticle and acrolein on AgPd FCC(111) surface shown in the illustration example. We measured the required DFT calls to fully relax the configurations and compared it with the built-in VASP quasi-Newton optimizer RMM-DIIS. We relaxed the configurations until the maximum force on the atoms is less than 0.05 eV/\(\AA\). The results are shown in Figure \ref{slab-nano-ads}. Active learning accelerates the relaxation process to different extents across these three systems. For the simpler case like the AuPd bare slab, the acceleration ratio is about 50\% compared to the pure VASP optimizer. For more complicated (i.e. lower symmetry and more atomic degrees of freedom) systems, the acceleration was more significant, reducing the number of DFT calls by more than 90\%. This result shows that active learning is suitable for relaxing more complicated structures. Once the NN has a reasonable representation of the potential energy surface of the target configurations by calling the first several DFT calculations, this surrogate model could be used to fine-tune the structure as a replacement of the DFT calls.

\begin{figure}[!htpb]
\centering
\includegraphics[width=.9\linewidth]{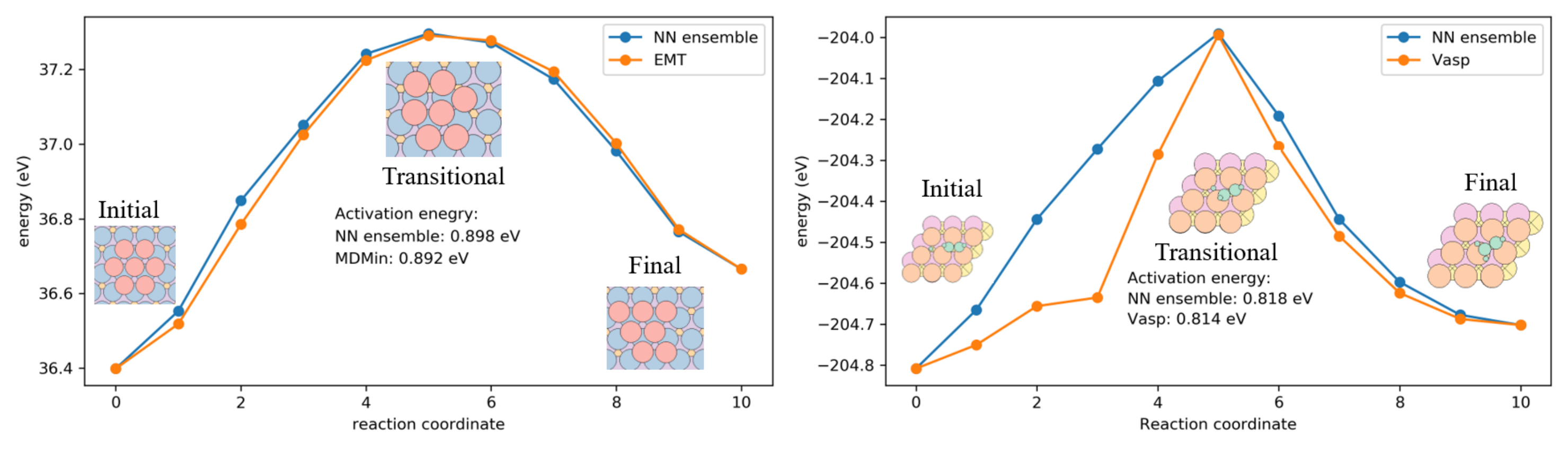}
\caption{\label{neb-result}Climbing NEB curves generated by NN ensemble and (a) EMT for Pt heptamer rearrangement (b) DFT for acetylene hydrogenation over Pd FCC(111) surface.}
\end{figure}

In addition to the local geometry optimization in the aforementioned cases, we also evaluated the NN ensemble based active learning method in two climbing image NEB (CINEB) examples: Pt heptamer rearrangement over Pt FCC(111) surface and acetylene hydrogenation over Pd FCC(111) surface.  We use an effective medium theory (EMT) calculator for the heptamer and DFT for the hydrogenation reaction. \cite{jacobsen-1996-semi-empir} We use EMT for heptamer because of the large size of the Pt slab. This example also shows that the NN ensemble method is not limited to DFT. We note that EMT is a relatively simpler potential than DFT, thus, we also include the acetylene hydrogenation with DFT as an example. The reaction curves generated by the NN ensemble with active learning and the corresponding VASP or EMT calculator are shown in Figure \ref{neb-result}. With the same initial and final state, the NN ensemble found practically the same transition state as VASP or EMT for these two system. The corresponding activation energies have 6 meV and 4 meV error compared to the one from EMT or DFT which is within convergence tolerance. The required DFT or EMT calls are much fewer than those without active learning as shown in Table \ref{neb-call}. In the case of acetylene hydrogenation, there are some mismatched energies between NN and VASP for the intermediate configurations except the transition state. This is caused by the intrinsic setting of the low scaling CINEB method based on active learning \cite{torres-2019-low-scalin}. Only  DFT data for the configuration with the highest energy is evaluated for the convergence criterion. This problem could be alleviated by modifying the convergence criterion to include the energy and forces of other images in the elastic band, such that all images in the band are fully relaxed instead of only considering the highest-energy configuration \cite{koistinen-2017-nudged-elast}. However, for the purpose of CINEB, the NN ensemble with active learning could accelerate the process to find the transition state by finding the configuration with the highest energy.

\begin{table}[!htpb]
\caption{\label{neb-call}EMT or DFT calls queried by NN emsemble with active learning, EMT with MDMin and VASP with built-in quasi newton optimizer for Pt heptamer rearrangement and acetylene hydrogenation.}
\centering
\begin{tabular}{lll}
\hline
 & Pt heptamer rearrangement & Acetylene Hydrogenation\\
 & (EMT) & (VASP)\\
\hline
Calculator & 596 calls & 1109 calls\\
\hline
NN ensemble with AL & 9 calls & 30 calls\\
\hline
\end{tabular}
\end{table}

\subsection{Limiting the training data to recent configurations for training efficiency}
\label{sec:orgeb15923}

\begin{figure}[!htpb]
\centering
\includegraphics[width=.9\linewidth]{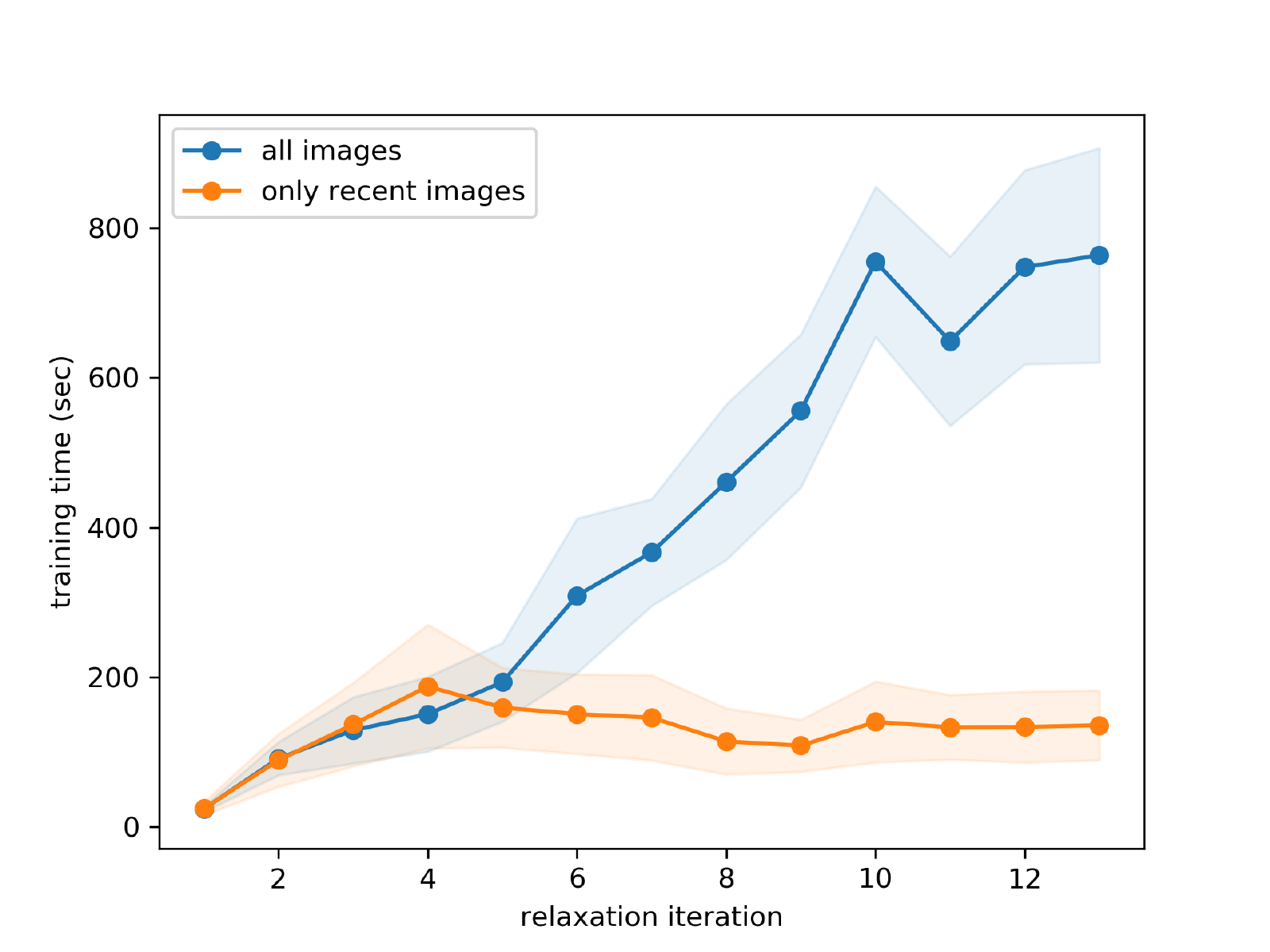}
\caption{\label{training-size}Time spent on the training process using a single NN with 2 layers and 50 neurons at each layer over iterations. The blue line shows the time for the model trained on all queried configurations while the orange line shows the time for training on the training set with fixed size. The experiment is repeated 10 times and the shaded area is the standard deviation for the 10 experiments. Time measured on 4 CPU cores.}
\end{figure}

With the active learning approach we add training data as the geometry optimizations proceed. This also adds (re)-training time which grows as the size of the training set. In the first few steps from scratch, this is not a problem since the training process could be completed quickly because of the small size of the training set. The time cost for training is  negligible compared to the DFT calculations. However, when the size of the training set grows large compared to the relaxation steps, the required time to train a model with high accuracy also scales up. Figure \ref{training-size} illustrates the training time for the NN over the active learning iterations. The initial training set consists of 13 different acrolein/AgPd configurations. At each iteration, uncertain configurations are added into the training set and the surrogate model is updated. The training cost time scales linearly with the size of the training set, which could be time consuming when the iterations increase.

\begin{figure}[!htpb]
\centering
\includegraphics[width=.9\linewidth]{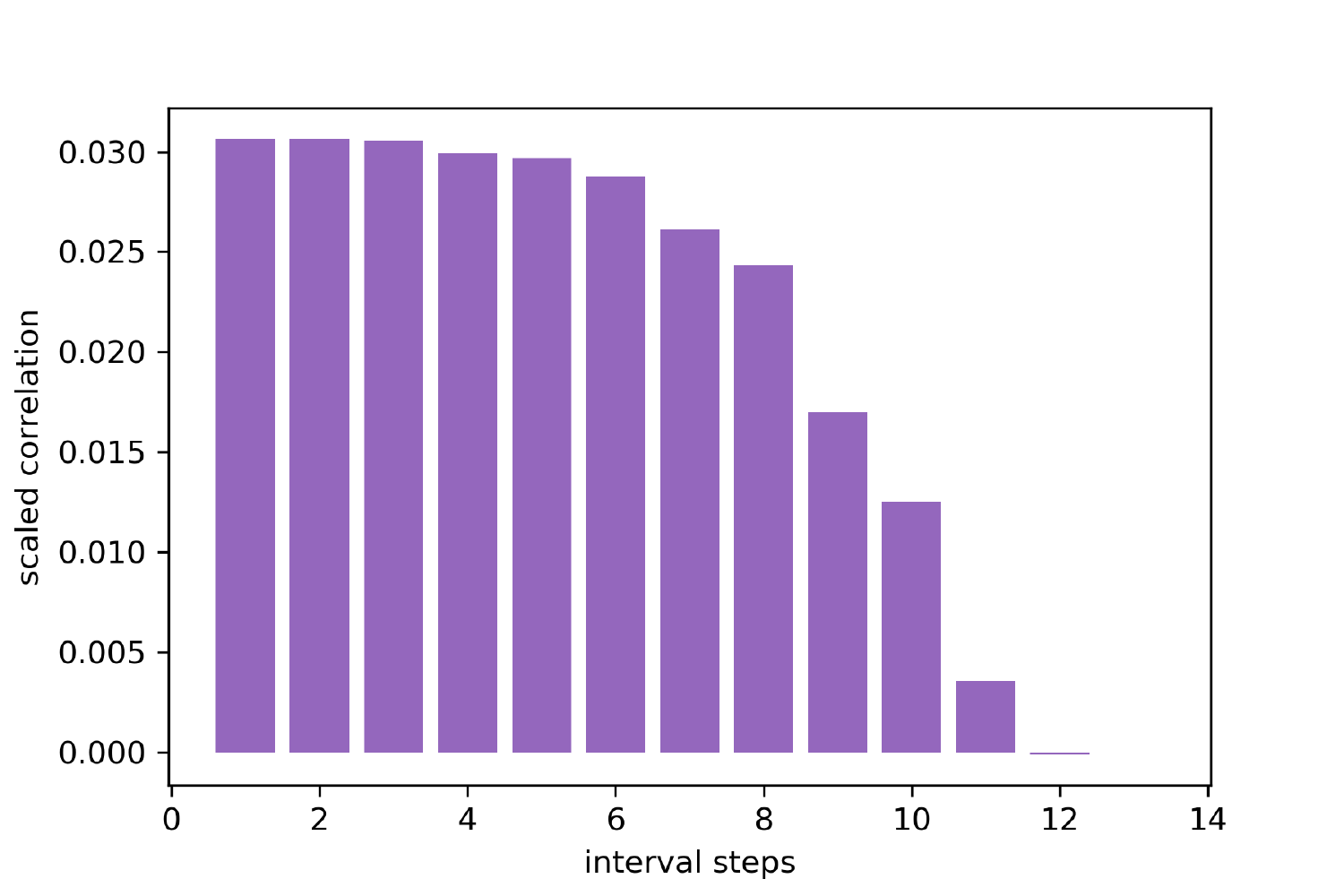}
\caption{\label{correlation}Scaled Pearson correlation coefficient between the intermediate configurations and the final relaxed configuration. The Pearson correlation is scaled by the base correlation between the initial configuration and the final configuration.}
\end{figure}

It is not always necessary to use all of the training data however. We found that the correlation (or similarity) between two configurations in the relaxation trajectories decreases as the number of steps between them increases. The correlation  between two configurations can be illustrated by averaging the Pearson correlations between corresponding atomic fingerprints in two configurations. There is usually reasonable similarity between the initial and final states (assuming a reasonable initial guess is used), so to highlight the change in similarity we subtracted the final state correlation from each configuration because the relaxation is local. The descending correlation shown in Figure \ref{correlation} for a relaxation trajectory suggests we may only need to focus on utilizing the configurations in the most recent steps to perform locally geometry relaxation.

As a result of Fig.  \ref{correlation} it appears in this system at least that after about five steps the new steps are decreasingly correlated with the initial steps. Therefore, if we only focus on recent steps (e.g. the five most recent steps) and only use these configurations to update the surrogate model, the training time could be controlled as almost constant as the active learning proceeds (see Figure \ref{training-size}). We note in this case that the training time is still small compared to the time required for a single DFT calculation which is about 1.5 hours for the Acrolein/AgPd unit cell with the VASP settings in this work. When the total training set continues to grow or there are fewer computational resources available for training, the local training set could be more preferable. We note that there are cheaper probabilistic models like GPR that could be used for small dataset. But given the growing size of the available data and the wide applications of deep learning models, a cheaper way to access the uncertainty estimation for deep learning models is valuable.

\section{Conclusion}
\label{sec:org8243325}

Active learning has demonstrated promising performance to accelerate the structure optimization in various applications. In this work, we illustrate that active learning with multiple configurations could achieve further acceleration compared to the active learning with single configuration by sharing the information across different configuration using a common NN ensemble. On the basis of that, we also provide three active learning modes for three scenarios with different amount of prior data. By integrating the prior data into the active learning framework, more calls to expensive energy and force calculators are saved. To explore the generalization ability of this method, we compared the number of required underlying energetic calculations between the active learning, built-in VASP quasi-Newton optimizer and BFGS in ASE in various local geometry optimization tasks. The results show that active learning reduces the amount of DFT or EMT calls by 50\% - 90\% based on different systems. From bare slabs to surfaces with adsorbates, the acceleration becomes more significant. In addition to the surface relaxation, we also applied this method to the climbing NEB for Pt heptamer rearrangement and acetylene hydrogenation. In these examples, the acceleration is even more apparent (\textasciitilde{}98\%) while keeping almost the same transition state with the underlying ground truth energy and force calculators. In conclusion, this work shows the potential of this NN ensemble based active learning method in various computational surface science and catalysis tasks.

\section{Supplementary Material}
\label{sec:org5306cc1}

See supplementary material for specific information about the code used in this work and instructions for accessing the datasets used in this work.

\begin{acknowledgments}
This material is based upon work supported by the U.S. Department of Energy, Office of Science, Office of Basic Energy Sciences, Catalysis program under Award Number DE-SC0018187. OAJN was supported under NSF DMREF Award CBET-1921946.
\end{acknowledgments}

\section{Data availability}
\label{sec:org432ec59}
The data that supports the findings of this study are available within the article [and its supplementary material].

\end{document}